\documentclass[conference]{IEEEtran}

\usepackage{amsmath}
\usepackage{amsfonts}%
\usepackage{amssymb}%
\usepackage{amsthm}
\usepackage{amssymb}
\usepackage{slashbox}
\usepackage{color,soul}
\usepackage[font={small,it}]{caption}
\usepackage{subfig}

\usepackage{float}
\usepackage{graphicx}

\numberwithin{equation}{section}

\newtheorem{theorem}{Theorem}[section]

\newtheorem{definition}[theorem]{Definition}

\begin{document}

\title{Unified Convex Optimization Approach to Super-Resolution Based on Localized Kernels}

\author{\IEEEauthorblockN{Tamir Bendory}
\IEEEauthorblockA{Electrical Engineering\\
Technion - Israel Institute of Technology
}

\and

\IEEEauthorblockN{Shai Dekel}
\IEEEauthorblockA{GE Global Research \\ School of Mathematical Sciences \\
Tel-Aviv University}

\and

\IEEEauthorblockA{Arie Feuer}
\IEEEauthorblockA{Electrical Engineering\\
Technion - Israel Institute of Technology }
}

\maketitle

\begin{abstract}
The problem of resolving the fine details of a signal from its coarse scale measurements  or, as it is commonly referred to in the literature, the super-resolution problem arises naturally in engineering and physics  in a variety of settings. We suggest a unified convex optimization approach for super-resolution. The key is the construction of an interpolating polynomial in the measurements space based on localized kernels.
We also show that the localized kernels act as the connecting thread to another wide-spread problem of stream of pulses.
\end{abstract}

\section{Introduction}\label{sec:intro}
Super-resolution is the task of estimating the fine details of a signal from its coarse scale measurements (see for instance \cite{greenspan2009super,park2003super,elad1997restoration}). This problem arises in many physical and engineering problems as sensing systems have a physical constraint on the highest resolution the system can
achieve. The aim of this paper is to suggest a unified convex optimization approach for super-resolution problems, based on the construction of interpolating polynomials and the existence of well-localized kernels. 

For simplicity, we consider signals of the form 
\begin{equation} \label{eq:x_sr}
x(t)=\sum_mc_m\delta_{t_m},\quad t\in {M},
\end{equation}
where $\delta_t$ is a Dirac measure, $M\subset\mathbb{R}^d,\thinspace d\geq 1,$ is a compact manifold and $T:=\left\{t_m\right\}$ is the signal's support. This model can be extended to piece-wise polynomials given additional information (e.g. boundary condition, signal's average). 

The information we have on the signal is its low spectral coefficients, namely its 'low-resolution' measurements. 
In this paper, we consider three concrete examples. Suppose $M$ is the $d$-dimensional torus. In this case, we assume that the measurements are given as 
\begin{equation*}
y=F_Nx,
\end{equation*}
where $F_N$ projects $x$ onto the space of trigonometric polynomials of degree $N$. That is, we have access solely to the low $2N+1$ low Fourier coefficients of $x$. This model reflects the fact that many measurement devices  have  limited resolution (e.g.  diffraction limits in microscopes) . This problem can be solved using parametric methods such as MUSIC, matrix pencil and ESPRIT \cite{stoica2005spectral,56027,32276,1143830}. However, the robustness of these methods is not well understood. We mention that asymptotic results regarding the robustness of these methods were derived in \cite{clergeot1989performance,stoica1991statistical} and significant steps towards a non-asymptotic behavior have been taken recently \cite{liao2014music,moitra2014threshold}.

Another closely-related example is the projection of $x$ onto the space of algebraic polynomials of degree $N$. In this case we choose $M$ as the manifold $[-1,1]^d$. This problem arises, for instance, in spectral methods to numerically solve partial differential equations (see e.g. \cite{canuto1987spectral,shen1994efficient}).

The third example is low-resolution measurements of signals lying on the bivariate sphere $\mathbb{S}^2$. In this case, we assume that a Dirac ensemble of the form (\ref{eq:x_sr}) lies on $\mathbb{S}^2$, and we have access only to its low $N$ spherical harmonics coefficients, which is the extension of Fourier analysis for signals on the sphere. This problem arises for instance in medical imaging \cite{tournier2004direct,deslauriers2012spherical}.

The rest of the paper is organized as follows. The following section describes the convex optimization framework and its main ingredient - the construction of interpolating polynomials. Section \ref{sec:construction} reveals the intimate relations between localized kernels and super-resolution and demonstrates it on the three examples. Later on, Section \ref{sec:sop} is devoted to the problem of recovery from stream of pulses and its connection to the super-resolution problem.  Section \ref{sec:numeric_exp} shows several numerical experiments for super-resolution on the sphere. Ultimately, in Section \ref{sec:conclusions} we draw some conclusions and suggest future extensions. 

\section{Convex Optimization Approach to Super-resolution}

In this paper we focus on a convex optimization approach for resolving signals from their low resolution measurements. We use the Total-Variation (TV) norm as a sparse-promoting regularization. In essence, the TV norm is the generalization of $\ell_1$ norm to the real line (for rigorous definition, see for instance Section 1 in  \cite{bendory2013Legendre} and \cite{rudin1986real}). For signals of the form (\ref{eq:x_sr}), we have $\Vert x\Vert_{TV}=\sum_m\vert c_m\vert$.

The main pillar of this framework is the following theorem \cite{bendory2013exact}:
\begin{theorem}\label{Th:dual} 
Let $x(t)=\sum_m c_m\delta_{t_m} , c_m\in\mathbb{R}$, where $T:=\{t_m\}\subseteq M$, and $M$ is a compact manifold in 
$\mathbb{R}^d$. Let $\Pi_D$ be a linear space of continuous functions of dimension $D$ in $M$. For any basis $\{P_k\}_{k=1}^D$ of $\Pi_D$, let $y_k = \langle x,P_k\rangle$ for all $1\leq k\leq D$. If for any signed sequence set $\{ u_m \}\in \{-1,1\}$ there exists $q\in \Pi_D$ such that
\begin{align}
q(t_m)&=u_m \,,\, \forall t_m\in T  ,\label{eq:dual1} \\
\vert q(t)\vert&<1 \,,\, \forall t\in M\backslash T, \label{eq:dual2}
\end{align}
then $x$ is the unique real Borel measure satisfying 
\begin{equation}
\min_{\tilde{x}\in\mathcal{M}(M) }\|\tilde{x}\|_{TV} \quad \mbox{subject to} \quad y_k = \langle \tilde{x},P_k\rangle  \,,\, 1\leq k\leq D,
\label{eq:TV_min_duality}
\end{equation}
 where $\mathcal{M}(M)$ is the space of signed Borel
measures on $M$. 
\end{theorem}

By taking $\Pi_D$ as the space spanned by the eigenfunctions of the low-resolution measurement operator, Theorem \ref{Th:dual} implies that the super-resolution problem can be reduced to a polynomial construction problem.  We mention that similar techniques, frequently referred to as \emph{dual certificate}, are widely used in many sparse recovery problems, see for instance \cite{candes2014mathematics,tang2012compressive,de2012exact}.

From the algorithmic perspective, it is interesting to notice that in all three cases the infinite-dimensional TV minimization problem (\ref{eq:TV_min_duality}) can be solved accurately (i.e. to any desired resolution) by algorithms with finite complexity which depends only on $N$. The recovery is performed by a three-stage algorithm involving the solution of a semi-definite program based on the dual problem of (\ref{eq:TV_min_duality}) (can be solved using off-the-shelf software), followed by root-finding and least square fitting \cite{candes2013towards,bendory2013Legendre,SR_sphere}. Alternatively, one can discretize the problem and solve standard $\ell_1$ minimization which  converges to the solution of (\ref{eq:TV_min_duality}) as the discretization becomes finer \cite{tang2013sparse}. As we discuss later, these solutions are robust to noisy measurements under a separation condition.

\section{The Construction of Interpolating Polynomials} \label{sec:construction}

As aforementioned, super-resolution problem can be reduced to the construction of an interpolating polynomial which lies in the space spanned by the eigenfunctions of the low-resolution measurement operator. The existence of such a polynomial relies on two interrelated pillars. The first is the separation condition defined as follows.

\begin{definition}\label{def:separation}
A set of points $T\subset M$ is said to satisfy the minimal separation condition with respect to the metric $d(\cdot,\cdot)$ if
\begin{equation*}
\Delta:=\min_{t_{i},t_{j}\in T,t_{i}\neq t_{j}} d\left( t_i,t_j\right) \geq\nu/N,
\end{equation*}
 where $\nu>0$ is a constant which does not depend on $N$.
\end{definition}  

Along this work we argue that the separation condition is a sufficient condition. However, we emphasize that few works prove that the separation is also necessary, and without minimal separation the recovery can not be stable in the presence of superfluous noise \cite{candes2013towards,SR_sphere,moitra2014threshold}.

The second pillar is the existence of well-localized kernels in the space spanned by the eigenfunction of the low-resolution measurement operator. To explain this statement, we will demonstrate it through our three prime examples. 

Consider a signal of the form of (\ref{eq:x_sr}) defined on the circle $[0,1]$ and its projection onto the space of trigonometric polynomials of degree $N$.  In the time domain, the projection is given as
\begin{equation*} \label{eq:Fourier}
y(t)=(x\ast D_N)(t),
\end{equation*}
where $D_N(t)=\sum_{k=-N}^Ne^{jkt}$ is the Dirichlet kernel of degree $N$.

The key for super-resolving the signal is the existence the Fejer kernel $\tilde{D}_N$, a well-localized smooth super-position of Dirichlet kernels. Equipped with the wrap-around metric $d\left( t_i,t_j\right)=\Vert t_i-t_j\Vert_\infty$, the authors of \cite{candes2013towards} showed that under the separation condition of Definition \ref{def:separation}, there exist coefficients $\{a_m\}$ and $\{b_m\}$ so that the polynomial 
\begin{equation*}
q(t)=\sum_m a_m\tilde{D}_N\left(t-t_m\right)+ b_m\tilde{D}_N^{(1)}\left(t-t_m\right),
\end{equation*}
satisfies (\ref{eq:dual1}) and (\ref{eq:dual2}).
Hence, the TV minimization (\ref{eq:TV_min_duality}) recovers $x$ exactly in the univariate case. They also showed that similar polynomials can be constructed for higher dimensional signals. In this manner, the existence of the Fejer kernel is crucial for the ability to resolve a signal on the d-dimensional torus under the separation condition. In consecutive papers, the existence of the interpolating polynomial also serves as the basis for proving that the TV minimization results in a robust and localized solution in the presence of noise \cite{candes2013super,fernandez2013support}. 

A similar result holds for the case in which the measurements are the projection of a signal of the form (\ref{eq:x_sr}) on $[-1,1]$ onto any basis spanning the space of algebraic polynomials of degree $N$. As was shown in \cite{bendory2013Legendre}, by considering the metric $d\left(t_i,t_j\right)=\left\vert\arccos \left(t_i\right) - \arccos \left(t_j\right)\right\vert$ one can construct the appropriate algebraic polynomial $q(t)$ under the separation condition of Definition \ref{def:separation} and hence the recovery through TV minimization is guaranteed. Observe that in this case, the minimal required separation is space-dependent and reduced near the edges to the order of $\mathcal{O}\left(N^{-3/2}\right)$. A consecutive paper showed that the recovery is also robust to noisy measurements \cite{de2014non}. These results hold for the bivariate case as well.

The same phenomenon occurs for signals on the bivariate sphere $\mathbb{S}^2$. Recall that spherical harmonics is the natural extension of Fourier analysis to signals on the sphere, and consider a Dirac ensemble on $\mathbb{S}^2$
\begin{equation} \label{eq:signal_sphere}
x(\xi)=\sum_m c_m\delta_{\mathbf{\xi}_m}, \quad\left\{\xi_m\right\}\in\Xi\subset \mathbb{S}^2.
\end{equation}

Let $Y_{n,k}(\xi), \thinspace  0\leq n\leq N, \thinspace -n\leq k\leq n$ be an orthonormal basis of  $V_N\left(\mathbb{S}^2\right)$, the space of spherical harmonics of degree $N$.
We assume that the information we have on the signal is its projection onto $V_N\left(\mathbb{S}^2\right)$
\begin{equation} \label{eq:low_res_sphere}
\hat{y}_{n,k}=\langle x, Y_{n,k}\rangle,\quad 0\leq n\leq N , \quad -n\leq k \leq n.
\end{equation} 

In the space domain, the projection onto $V_N\left(\mathbb{S}^2\right)$  can be written as a spherical convolution, 
\begin{equation*}
y(\xi)=\int_{\mathbb{S}^2}x(\eta)K_N(\xi\cdot\eta)d\mathbb{S}^2(\eta), 
\end{equation*}
where by the addition formula \cite{atkinson2012spherical}
\begin{equation*}
K_N(\xi\cdot\eta)=\sum_{k=-N}^NY_{n,k}(\xi)\overline{Y}_{n,k}(\eta)=\frac{2N+1}{4\pi}P_{N,3}(\xi\cdot\eta),
\end{equation*}
where $\overline{Y}_{n,k}$ is the conjugate of ${Y}_{n,k}$, and $P_{N,3}(x)$ is the univariate ultraspherical Gegenbauer polynomial of order 3 and degree $N$. 

In \cite{narcowich2006decomposition}, it was shown that a smooth super-position of the kernel $K_N(\xi\cdot\eta)$ results in a well-localized kernel in the space of spherical harmonics of degree $N$. We denote the localized kernel as $F_N(\xi\cdot\eta)$, and by $D_{\xi,\ell},\thinspace \ell=1,2$ the partial rotational derivatives at $\xi$. Leveraging the localization of $F_N$, it is known \cite{bendory2013exact} that there exist coefficients $ \{\alpha_m\}$, $\{\beta_m\}$ and $ \{\gamma_m\}$ so that a polynomial of the form 
\begin{equation*}
\begin{split}
q(\xi)&=\sum_m \alpha_m F_N(\xi\cdot\xi_m)+\beta_m D_{\xi_m,1}F_N(\xi,\xi_m) \\ &+\gamma_m D_{\xi_m,2}F_N(\xi,\xi_m),
\end{split}
\end{equation*}
fulfils the requirements of Theorem {\ref{Th:dual} under the separation condition with the natural distance on the sphere $d\left(\xi_i,\xi_j\right)=\arccos\left(\xi_i\cdot\xi_j\right)$. Accordingly, we present the following theorem:
\begin{theorem}
\label{Th:sphere} \cite{bendory2013exact} Let $\Xi=\{\xi_{m}\}$ be the support of a signed
measure x of the form $(\ref{eq:signal_sphere})$. Let $\{Y_{n,k}\}_{n=0}^{N}$
be any spherical harmonics basis for $V_{N}(\mathbb{S}^{2})$ and
let $\hat{y}_{n,k}=\langle x,Y_{n,k}\rangle$, $0\le n\le N$, $-n\le k\le n$.
If $\Xi$ satisfies the separation condition of Definition \ref{def:separation} with the metric $d\left(\xi_i,\xi_j\right)=\arccos\left(\xi_i\cdot\xi_j\right)$,
then $x$ is the unique solution of
\begin{equation} \label{eq:TV_sphere}
\begin{split}
&\min_{\tilde{x}\in\mathcal{M}(\mathbb{S}^{2})}\|\tilde{x}\|_{TV}\quad\mbox{subject to}\quad \hat{y}_{n,k}=\langle \tilde{x},Y_{n,k}\rangle \\ 
& 0\leq n\leq N,\thinspace  -n \leq k\leq n,
\end{split}
\end{equation}
 where $\mathcal{M}(\mathbb{S}^{2})$ is the space of signed Borel
measures on $\mathbb{S}^{2}$. 
\end{theorem}
It has been shown in \cite{SR_sphere} (for a discrete version of (\ref{eq:TV_sphere})) that the recovery error in the presence of noise is proportional to the noise level.

We conclude this section with an interesting observation regrading non-negative signals (i.e. $c_m>0$). In this case, a sufficient condition for signal recovery is the existence of interpolating polynomial as in Theorem \ref{Th:dual} but the constraint in (\ref{eq:dual1}) is replaced by a weaker constraint of $q(t_m)=1$ for all $t_m\in T$ (see  Theorem 5.1 in \cite{bendory2013exact}). Consider the case of the projection of signals on $[0,1]$ onto the space of trigonometric polynomials of degree $N$. In this case for any $s\leq N$ a polynomial of the form 
\begin{equation*}
q(t)=1-2^{-(s+1)}\prod_{m=1}^s\left(1-\cos\left(t-t_m\right)\right),
\end{equation*}
is a trigonometric polynomial of degree $N$, $q(t_m)=1$ for all $t_m\in T$ and $\vert q(t)\vert<1$ otherwise. Consequently, a \emph{clustered} $N$-sparse signal with non-negative coefficient can be recovered exactly. The same observation is noted for the other two cases with similar constructions (see \cite{bendory2013exact}).

\section{Super-resolution and Stream of Pulses} \label{sec:sop}
In the previous section we stressed that the existence of well-localized kernels is crucial for resolving signals. The localized kernels bind the super-resolution problem to the problem of stream of pulses, namely recovery of the delays $T:=\left \{t_m \right \}$ and weights $\left \{c_m\right \}$ from the measurements 
\begin{equation*} \label{eq:y}
y(t)=\sum_mc_m K_\sigma\left(t-t_m\right), \quad c_m,t\in\mathbb{R},
\end{equation*}
where $K(t)$ is a pulse shape (kernel) and $K_\sigma(t):=K(t/\sigma)$ for some scaling parameter $\sigma>0$.
An alternative representation of the problem is as $y(t)= \left( x \ast K_\sigma\right)(t)$  where 
\begin{equation*} \label{eq:x}
x(t)=\sum_mc_m\delta_{t_m},\quad T:=\left\{t_m \right\},
\end{equation*}
and $\delta_t$ is a Dirac measure. 

In a manner similar to Theorem \ref{Th:dual}, the recovery problem can be reduced to construction of a special interpolating function, as follows:
\begin{theorem}
\label{th:sop}\cite{SOP} Let $x(t)=\sum_{m}c_{m}\delta_{t_{m}}$, \textup{$c_{m}%
\in\mathbb{R}$} where $T:=\{t_{m}\}\subseteq\mathbb{R}$, and let
$y(t)=\int_{\mathbb{R}}K(t-s)dx\left(  s\right)  $ for a $L$
times differentiable kernel $K(t)$. If for any set $\{u_{m}\}\in\{-1,1\}$ there exists a function of the form
\begin{equation*}
q(t)=\int_{\mathbb{R}}\sum_{\ell=0}^{L}K^{(\ell)}(s-t)d\mu_{\ell}\left(
s\right)  ,\label{7}%
\end{equation*}
for some measures \textup{$\left\{  \mu_{\ell}\left(  t\right)  \right\}
_{\ell=0}^{L}$}, satisfying
\begin{align*}
q(t_{m}) &  =u_{m}\,,\,\forall t_{m}\in T,\\
|q(t)| &  <1\,,\,\forall t\in\mathbb{R}\backslash T,%
\end{align*}
then $x$ is the unique real Borel measure solving
\begin{equation}
\begin{split}
&\min_{\tilde{x}\in\mathcal{M\left(  \mathbb{R}\right)  }}\Vert\tilde{x}
\Vert_{TV}\quad\mbox{subject to} \\ & y(t)=\int_{\mathbb{R}}K(t-s)d\tilde
{x}\left(  s\right)  . \label{eq:TVminSOP}%
\end{split}
\end{equation}
\end{theorem} 

We note that here too, as in the super-resolution problems, the existence of $q(t)$ relies on separation and localization. If the support $T$ satisfies the separation condition with the metric $d(t_i,t_j)=\vert t_i-t_j\vert$ and $\sigma=1/N$, there exist coefficients $\{a_m\}$ and $\{b_m\}$ so that 
\begin{equation*} \label{eq:q}
q(t)=\sum_m a_mK_\sigma\left(t-t_m\right)+ b_mK_\sigma^{(1)}\left(t-t_m\right),
\end{equation*}
satisfies the requirements of Theorem \ref{th:sop}, and thus enabling perfect recovery. This holds if the kernel $K(t)$ is localized. More precisely, the kernel should satisfy the definition of \emph{admissible} kernel as follows:
\begin{definition}
\label{Definition2} A kernel $K$ is {admissible} if it
has the following properties:

\begin{enumerate}
\item $K\in \mathcal{C}^3\left(\mathbb{R}\right)  $, is real and even.

\item \underline{Global property:} There exist constants $C_{\ell}>0, \ell=0,1,2,3$ such
that $\left\vert K^{\left(  \ell\right)  }\left(  t\right)  \right\vert
\leq\ C_{\ell} / \left( {1+t^{2}} \right)$.

\item \underline{Local property:} There exist constants $\varepsilon,\beta>0$ such that 
\begin{enumerate}
\item  $K(t)>0$ for all $\vert t\vert \leq \varepsilon$ and $K(t)  < K(\varepsilon)$ for all $\vert t\vert>\varepsilon$,
\item  $K^{\left(2\right)  }\left( t\right)  <-\beta$  for all $\vert t\vert \leq \varepsilon$.
\end{enumerate}
\end{enumerate}
\end{definition}

Here too, the existence of the interpolating function guarantees a robust and localized recovery \cite{SOP,SOP_US}. Interestingly, in the non-negative case (i.e. $c_m>0$) the separation is unnecessary. In the presence of noise, the recovery error is proportional to the noise level, and depends on the  number of spikes within any resolution cell of size $\nu\sigma$ \cite{SOP_positive}. 

\section{Numerical Experiments on the Sphere}\label{sec:numeric_exp}
To demonstrate our results we chose to use the less familiar example where we consider the recovery of a Dirac ensemble on the sphere (\ref{eq:signal_sphere}) from its low-resolution measurements (\ref{eq:low_res_sphere}). In this case, the TV minimization (\ref{eq:TV_sphere}) can be solved to any desired accuracy by a three-stages algorithm consists of  semi-definite optimization problem with $\mathcal{O}\left(N^4\right)$ variables, followed by root-finding and least square fitting. In case that the measurements are contaminated with bounded noise, it has been shown that the recovery error is proportional to the noise level (in the discrete setting) \cite{SR_sphere}.

The following experiments were conducted in Matlab using CVX
\cite{cvx}, which is the standard modelling system for convex optimization.
The signals were generated in the following two stages:
\begin{itemize}
\item Random locations on the sphere were drawn uniformly, sequentially added to the signal's
support, while maintaining the separation condition of Definition \ref{def:separation}. 
\item Once the support was determined, the amplitudes were drawn randomly
from an iid normal distribution with standard deviation of $SD=10$.
\end{itemize}

Figure \ref{fig:separation} presents a numerical estimation of the separation constant $\nu$ (see Definition \ref{def:separation}). As can be seen, a separation constant of $2\pi$ seems to be sufficient in the noise-free setting. This separation coincides with the spatial resolution of the projection of $x$ onto $V_N$ \cite{rafaely2004plane}. 

\begin{figure}
\includegraphics[scale=0.3]{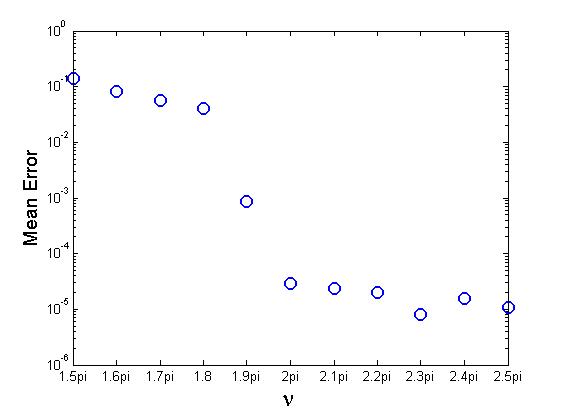} 
\protect\caption{The mean recoery error (in log scale) as a function of $\nu$ over 20 simulations. To be clear, by error we merely mean the distance on the sphere between the true and the estimated support. \label{fig:separation}}
\end{figure}

Figure \ref{fig:result1} presents an example for super-resolution on the sphere. The signal in Figure \ref{fig:result1a} is the projection of the signal onto $V_{10}$, whereas Figure \ref{fig:result1b} presents the recovered signal. The maximal and average recovery errors for several values of $N$ are presented in Table \ref{tab:1}. 

In the noisy setting, we added an iid noise with normal distribution and zero mean. Figure \ref{fig:result_noise} presents the mean recovery error as a function of the noise standard deviation. We note that the error increases moderately with the standard deviation. 

\begin{figure}

\subfloat[The low resolution measurements. \label{fig:result1a}]{
\includegraphics[scale=0.4]{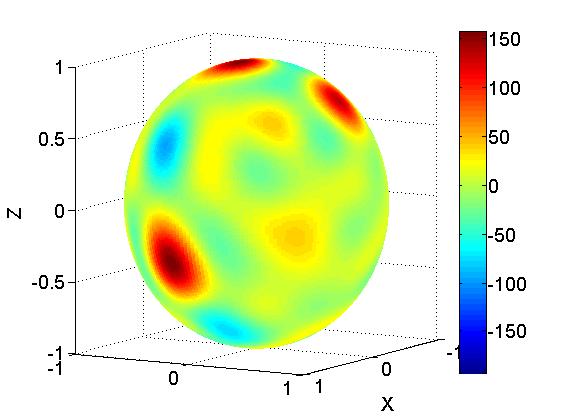}
}

\subfloat[The recovered signal $f$.\label{fig:result1b}]{
\includegraphics[scale=0.4]{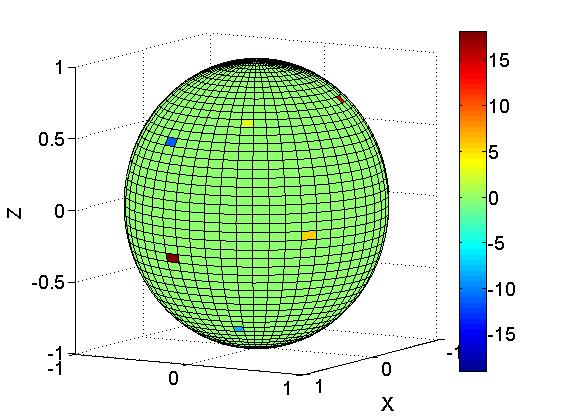}
}
\caption{Super-resolution on the sphere for $N=10$.
The signal is presented on a grid for visualization only.\label{fig:result1}}
\end{figure}

\begin{table}
\begin{centering}
\begin{tabular}{|c|c|c|c|}
\hline 
\multicolumn{1}{|c|}{N} & 5 & 8 & 10\tabularnewline
\hline 
\hline 
Average error & $8.1267\times10^{-5}$ & $8.1826\times10^{-5}$ & $9.0404\times10^{-5}$\tabularnewline
\hline 
Max error & $2.163\times10^{-4}$ & $1.9\times10^{-3}$ & $3.3\times10^{-3}$\tabularnewline
\hline 
\end{tabular}
\par\end{centering}

\protect\caption{The localization error for $N=5,8,10.$ For
each value of $N$, 10 experiments were conducted. To be clear, by error we merely mean the distance on the sphere between the true and the estimated support.\label{tab:1}}
\end{table}

\begin{figure}
\includegraphics[scale=0.3]{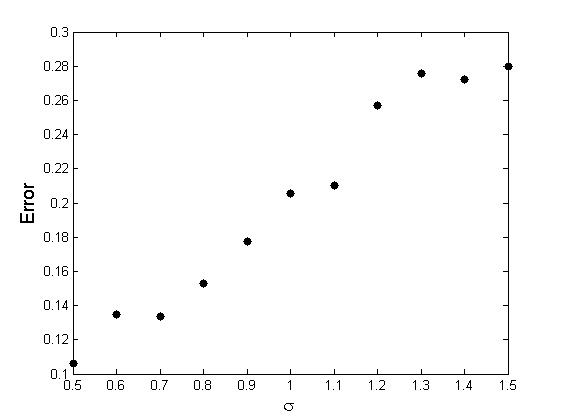}
\caption{ 10 experiments were conducted with $N=8$ for each value of standard deviation. The figure presents the average recovery error as a function of the noise standard deviation. 
 \label{fig:result_noise}}
\end{figure}

\section{Discussion}\label{sec:conclusions}
In this paper, we have presented a general framework for resolving robustly signals in various settings and geometries using convex optimization. Localized kernels play a crucial role in the process. An important question is how general is this framework. That is to say, under which settings we can expect to resolve signal robustly  using convex optimization. This topic is under ongoing research. 

We also showed that the localization principle relates super-resolution to other problems such as the recovery from stream of pulses. 
Recently, new results on the localization of the solution of (\ref{eq:TVminSOP}) have been developed and tested on real ultrasound data \cite{SOP_US}. It is of a great interest to find more applications where similar techniques could be applied (for recent related works, see for instance \cite{heckel2014super,aubel2014super}).

From the algorithmic perspective, the aforementioned super-resolution problems can be recast as finite dimensional problems which involve the solution of a semi-definite program, root-finding and least square fitting. It will be interesting to look for the relations of these algorithms with 'traditional' parametric methods such as MUSIC which  also rely on root-finding.

\bibliographystyle{IEEEtran}
\bibliography{IEEEabrv,bib}
\end{document}